# REAL-TIME NEARFIELD ACOUSTIC HOLOGRAPHY: IMPLEMENTATION OF THE DIRECT AND INVERSE IMPULSE RESPONSES IN THE TIME-WAVENUMBER DOMAIN

Jean-Hugh Thomas[1,2], Vincent Grulier[1], Sébastien Paillasseur[1,2] and Jean-Claude Pascal[1,2]

[1]LAUM, CNRS, Université du Maine, avenue 0. Messiaen, 72085 Le Mans, France
[2]ENSIM, rue Aristote, 72085 Le Mans, France
jean-hugh.thomas@univ-lemans.fr

**Abstract**

The aim of the study is to demonstrate that some methods are more relevant for implementing the Real-Time Nearfield Acoustic Holography than others. First by focusing on the forward propagation problem, different approaches are compared to build the impulse response to be used. One of them in particular is computed by an inverse Fourier transform applied to the theoretical transfer function for propagation in the frequency-wavenumber domain. Others are obtained by directly sampling an analytical impulse response in the time-wavenumber domain or by additional low-pass filtering. To estimate the performance of each impulse response, a simulation test involving several monopoles excited by non stationary signals is presented and some features are proposed to assess the accuracy of the temporal signals resulting from reconstruction processing on a forward plane. Then several inverse impulse responses used to solve the inverse problem, which consists in back propagating the acoustic signals acquired by the microphone array, are built directly from a transfer function or by using Wiener inverse filtering from the direct impulse responses obtained for the direct problem. Another simulation test is performed to compare the signals reconstructed on the source plane. The same indicators as for the propagation study are used to highlight the differences between the methods tested for solving the Holography inverse problem.

## 1. INTRODUCTION

Real-time Nearfield Acoustic Holography (RT-NAH) is an interesting method for characterizing and locating non stationary sources [1]. By operating in the time-wavenumber domain, using a convolution between each point of the wavenumber spectrum and an inverse impulse response, it is then possible to reconstruct continuously the time signals emitted from any point of the source plane facing the microphones of an array in the nearfield of the sources. This is the originality of RT-NAH in the field of acoustic Holography methods dedicated to the study of non stationary sources [2], [3], [4]. By considering the geometry of the problem (see figure 1), the time-dependent wavenumber spectrum $P(k_x, k_y, z_F, t)$ in a forward plane $z = z_F$ can be obtained by convolving each component of the time-dependent wavenumber spectrum $P(k_x, k_y, z_A, t)$ acquired in a measurement plane $z = z_A$ with an impulse response $h(k_x, k_y, z_F - z_A, t)$ in the time-wavenumber domain

$$P(k_x,k_y,z_F,t) = P(k_x,k_y,z_A,t) * h(k_x,k_y,z_F - z_A,t). \tag{1}$$

By using the following notations for the propagation distance $\Delta z = z_F - z_A$, the wavenumber $k_r = \sqrt{k_x^2 + k_y^2}$, the propagation delay $\tau = \Delta z / c$, and the transition pulsation $\Omega_r = c k_r$, the impulse response $h(k_x, k_y, \Delta z, t)$ of eq. (1) can be written [5], [6]

$$h(\Omega_r, \tau, t) = \delta(t - \tau) - \tau \Omega_r^2 \frac{J_1(\Omega_r \sqrt{t^2 - \tau^2})}{\Omega_r \sqrt{t^2 - \tau^2}} \Gamma(t - \tau). \tag{2}$$

$\delta(t)$ denotes the Dirac distribution and $\Gamma(t)$ is the Heaviside function.

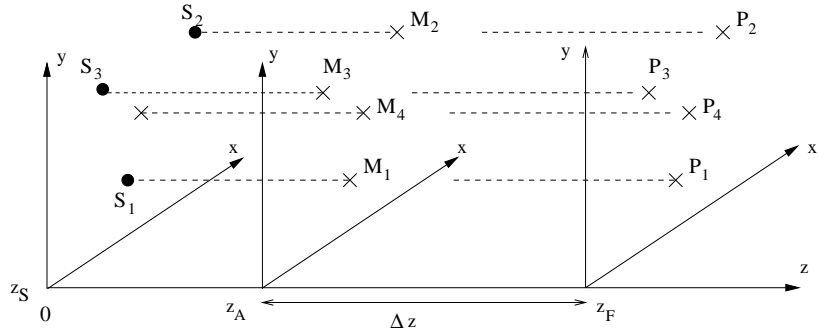

Figure 1. Geometry of interest. The direct problem consists in measuring the pressure field in the plane $z = z_A$ in order to compute the forward pressure in the plane $z = z_F$. For the inverse problem, the measurements are done from the plane $z = z_F$, the reconstructed pressure is computed in the plane $z = z_A$. Three monopoles $S_1$, $S_2$, and $S_3$ are simulated from the source plane $z = z_S$.

## 1. INTERPRETATION IN THE FREQUENCY DOMAIN

The frequency response $H(\Omega_r, \tau, f)$ is the Fourier transform with respect to time of the impulse response $h(\Omega_r, \tau, t)$. It can also be computed by applying a Fourier transform to Eq. (1). The equation obtained is then

$$P(k_x, k_y, z_F, \omega) = P(k_x, k_y, z_A, \omega) H(\Omega_r, \tau, \omega). \tag{3}$$

It reminds the relationships between the known pressure field on a plane $z = z_A$ and the pressure on any other plane $z = z_F$ when the studied stationary acoustic sources are confined on the half plane $z \leq z_S$ ($z = z_S$ is the source plane) [7]

$$P(k_x, k_y, z_F, \omega) = P(k_x, k_y, z_A, \omega) G_P(k_r, \Delta z, \omega), \tag{4}$$

where the propagator $G_P$ is defined by

$$G_P(k_r, \Delta z, \omega) = e^{-j k_z \Delta z} = \begin{cases} e^{-j \Delta z \sqrt{\left(\frac{\omega}{c}\right)^2 - k_r^2}} & \text{for } \omega/c \geq k_r \\ e^{-\Delta z \sqrt{k_r^2 - \left(\frac{\omega}{c}\right)^2}} & \text{for } \omega/c < k_r, \end{cases} \qquad (5)$$

where $c$ is the sound celerity. By using Eqs (3) to (5), the frequency response $H(\Omega_r, \tau, \omega)$ can be written

$$H(\Omega_r, \tau, \omega) = G_P(k_r, \Delta z, \omega) = \begin{cases} e^{-j \tau \sqrt{\omega^2 - \Omega_r^2}} & \text{for } \omega \geq \Omega_r \\ e^{-\tau \sqrt{\Omega_r^2 - \omega^2}} & \text{for } \omega < \Omega_r \end{cases} \qquad (6)$$

The modulus and the phase of the frequency response are represented with a dotted line in figure 2 for a transition frequency $f_r = \Omega_r/2\pi = 2000\,\text{Hz}$ and $\Delta z = 0.1075\,\text{m}$. The observation of this figure can make the reader understand the choice of the name "transition" for the specific frequency $f_r$. Indeed, $f_r$ is the frequency that separates two kinds of propagation for the acoustic waves: propagative waves for $f \geq f_r$ and exponentially decaying sound fields for $f < f_r$. The non stationary signal in the time-wavenumber domain $P(k_x, k_y, z_A, t)$, which is the time evolving pressure in the plane $z = z_A$ at the point $k_r$ of the wavenumber spectrum, will show its frequency components above the transition frequency propagate as propagative waves and its frequency components below $f_r$ decay exponentially.

## 2. PROCESSING FOR PROVIDING AN OPERATIONAL IMPULSE RESPONSE

Two approaches are considered here. The first is based on the analytical formulation of the impulse response $h(\Omega_r, \tau, t)$ given in Eq. (2) in the time-wavenumber domain. The second starts from the theoretical frequency response $H(\Omega_r, \tau, \omega)$ in Eq. (6) and by using an inverse Fourier transform, yields the impulse response.

### 2.1 Processing from the analytical response

For this case, let us consider the following equation derived from Eq. (2) giving the impulse response

$$h(\Omega_r, \tau, t) = \delta(t - \tau) - g(\Omega_r, \tau, t), \qquad (7)$$

where

$$g(\Omega_r, \tau, t) = \tau \Omega_r^2 \frac{J_1(\Omega_r \sqrt{t^2 - \tau^2})}{\Omega_r \sqrt{t^2 - \tau^2}} \Gamma(t - \tau). \qquad (8)$$

As sampling the impulse response even with a relatively high rate may lead to distorsions in the transfer function (see figure 2), direct sampling is replaced by average sampling. Instead of considering $g[n]$, the sampling value of $g(t)$ at the time $t = n\Delta t$, the mean value $\bar{g}[n]$ is computed into an interval $\Delta t$ centered at $t = n\Delta t$

$$\bar{g}[n] = \frac{1}{\Delta t} \int_{n\Delta t - \Delta t/2}^{n\Delta t + \Delta t/2} g(t) \, dt. \tag{9}$$

The integral in Eq. (9) will be approximated by the trapezoidal formula.

Another modification is used to overcome the problem of the impulse response whose transfer function is not band limited. The process consists first in increasing the sampling rate of the impulse response by a factor $D$ so that the new sampling frequency is $f'_e = D f_e$. The response contains $D N$ samples. Then the upsampled response is filtered using a low-pass filter. Finally, the filtered impulse response is downsampling by the factor $1/D$ to ensure that the final sampling frequency $f'_e/D$ matches $f_e$, that of the acoustic signals acquired. The number of samples of the resulting impulse response is $N$.

Two low-pass filters have been experimented, one with an infinite impulse response (IIR), a Chebyshev filter, the other one with a finite impulse response (FIR) given by associating a cardinal sine with a Kaiser-Bessel window. The impulse response of the FIR filter is

$$w(t) = \tau \Omega_r^2 \frac{I_0(\beta\sqrt{1-(2t/T)^2})}{I_0(\beta)} \frac{\sin(\alpha \pi t f_e)}{\pi t}. \tag{10}$$

$I_0$ is the modified Bessel function of the first kind and order 0. $T$ is the duration of the Kaiser-Bessel window. $\alpha$ is linked to the cutoff frequency $f_c$ of the low-pass filter and $\beta$ is a parameter which sets the sidelobes of the Kaiser-Bessel window.

Two ways of implementing the convolution between the impulse response and the low-pass filter can be considered. First, the filtered response $g_f(t)$ can be provided using a discrete sum as

$$g_f[n] = \sum_m w[m] g[n-m]. \tag{11}$$

But it can also be computed using a numerical approximation of the following integral given by the trapezoidal method

$$g_f(t) = \int_{t-T/2}^{t+T/2} g(\theta) w(t-\theta) \, d\theta. \tag{12}$$

**2.2 Processing from the theoretical frequency response**

The Fourier transform of Eq. (7) yields

$$H(\Omega_r, \tau, \omega) = e^{-j\omega\tau} - G(\Omega_r, \tau, \omega). \tag{13}$$

Since $H(\Omega_r, \tau, \omega)$ is analytically defined in Eq. (7), so it is for the transfer function $G(\Omega_r, \tau, \omega)$ whose both theoretical modulus and phase are easily deducted from Eq. (6). By applying an inverse Fourier transform either to $H(\Omega_r,\tau,\omega)$ in Eq. (6) or to $G(\Omega_r,\tau,\omega)$ in Eq. (13), the impulse response $h(\Omega_r,\tau,t)$ or $g(\Omega_r,\tau,t)$ is obtained. From Eq. (7), both approaches provide finally the same impulse response $h(\Omega_r,\tau,t)$.

**2.3 Comparisons between transfer functions resulting from processing**

The aim of this part is to compare several transfer functions $G(\Omega_r,\tau,\omega)$ resulting from different processing in order to conclude on their relevance for resolving the source radiating problem. Three treatments are applied to the theoretical function $g(\Omega_r,\tau,t)$ in Eq. (8). They are summarized in the following

- Average method: $g(\Omega_r,\tau,t)$ is average sampled according to Eq. (9),

- Chebyshev filtering: $g(\Omega_r,\tau,t)$ is low-pass filtered using a Chebyshev filter with a cutoff frequency $f_c = 6400$ Hz. It is achieved by upsampling $g(\Omega_r,\tau,t)$ by the factor $D=8$, using the low-pass filter and then downsampling the resulting response by the factor $1/D$.

- Numerical Kaiser filtering: $g(\Omega_r,\tau,t)$ is low-pass filtered using a Kaiser-Bessel filter with a cutoff frequency $f_c = 6640$ Hz. An upsampling factor of $D=2$ is used and the integral in Eq. (12) is numerically computed using the trapezoidal method.

For all cases, $g(\Omega_r,\tau,t)$ is initially sampled with the sampling frequency $f_e = 16000$ Hz giving 256 samples. The propagation distance and the transition frequency are set to $\Delta z = 0.1075$ m and $f_r = 2000$ Hz. Figure 2 highlights the transfer functions $H(\Omega_r,\tau,f)$ (modulus and phase) for the three different processing. The frequency responses are obtained by applying a Fourier transform to the sampled response $g(\Omega_r,\tau,t)$ before using Eq. (13).

By comparing in figure 2 the three transfer functions to the one obtained by directly operating a Fourier transform on the sampled response even though the sampling frequency is higher (64000 Hz instead of 16000 Hz), it seems evident that the transfer functions provided by processing $g(\Omega_r,\tau,t)$ are more relevant. In addition, filtering $g(\Omega_r,\tau,t)$ in order to limit its frequency band is advantageous. The use of average sampling with no filter is less effective than filtering in particular in the frequency area of propagative components. The use of a FIR filter with a Kaiser-Bessel window seems more accurate than the use of a Chebyshev filter especially for the phase. The most accurate transfer function in figure 2 is obtained by numerically computing the integral of convolution involving the Kaiser-Bessel filter.

It is of course true that the best matching between the theoretical and the computed transfer functions occurs for the method based on the inverse Fourier transform, which is evident as the starting point of the approach, called here Fourier method, is precisely the theoretical frequency response.

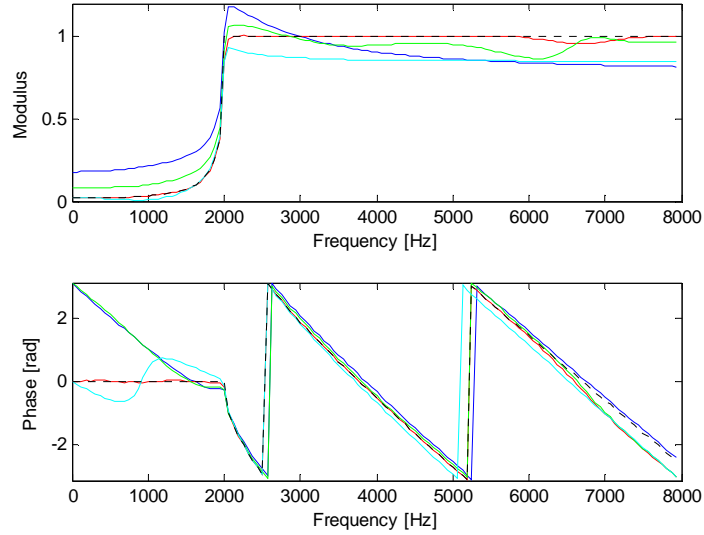

Figure 2. Modulus and phase of the transfer functions $H(\Omega_r, \tau, f)$ computed from the direct method (blue), the average method (cyan), the Chebyshev filtering (green), the numerical Kaiser method (red) and from equation (7) (- - - -). The transition frequency is $f_r = 2000\,\text{Hz}$, the sampling frequency is $f_e = 16000\,\text{Hz}$ except for the direct method ($f_e = 64000\,\text{Hz}$) and $\Delta z = 0.1075\,\text{m}$.

## 3. NUMERICAL RESULTS

### 3.1 Setup

The source plane is composed of three monopoles at the positions $S_1$ (0.3125m, 0.375m, 0m), $S_2$ (0.75m, 0.75m, 0m), and $S_3$ (0.25m, 0.75m, 0m). Both monopoles $S_1$ and $S_2$ generate a signal with a linear frequency modulation and a gaussian amplitude modulation while monopole $M_3$ radiates a Morlet wavelet whose expression is $s(t) = \cos(2\pi f t)\, e^{-t^2/2}$. Thus the sources are non stationary. The simulation of the acquisition is done using a 17×17 microphone array located in the measurement plane $z = z_A = 0.05\,\text{m}$. The step size in both $x$ and $y$ directions is $\Delta L = 0.0625\,\text{m}$, providing an overall scan dimension of $1.0 \times 1.0\,\text{m}^2$. The propagation distance is $\Delta z = 0.1075\,\text{m}$. Thus, the forward plane is located at $z_F = 0.1575\,\text{m}$ as shown in figure 1. The emitted signals are sampled at a frequency rate $f_e = 16000\,\text{Hz}$ providing 256 samples.

The first goal of this study is to reconstruct the time evolving pressure field at each point of the forward plane in front of the square grid of 17×17 virtual microphones using Eq. (1). Five different impulse responses $h(\Omega_r, \tau, t)$ in the time-wavenumber domain are tested. They are computed from the Chebyshev method (C), the numerical Kaiser method (K), the Fourier method (F), the average method (A) and the direct method (D) for which the impulse response $h(\Omega_r, \tau, t)$ is provided by directly sampling $g(\Omega_r, \tau, t)$ in Eq.(8) at $f_e = 64000\,\text{Hz}$. The second goal is to solve the inverse problem: assuming that the measurements are done from the plane $z = z_F$, the pressure field has to be back propagated to the plane $z = z_A$ using

$$P(k_x, k_y, z_A, t) = P(k_x, k_y, z_F, t) * h^{-1}(k_x, k_y, z_F - z_A, t). \qquad (14)$$

Two approaches are used to compute the inverse impulse response $h^{-1}(k_x, k_y, z_F - z_A, t)$.

The first consists in using Wiener adaptive filtering to invert the impulse responses obtained in the previous study [1], [8]. The second, denoted I, directly operates an inverse Fourier transform to the analytical inverse transfer function $1/H(\Omega_r,\tau,f)$ defined in Eq. (6).

### 3.2 Indicators

In order to comment the results obtained objectively, two temporal indicators $T_1$ and $T_2$ are proposed. They are based on the reconstructed signals $p(x,y,z_R,t)$ but also on the simulated signals $p_r(x,y,z_R,t)$ directly propagated on the reconstructed plane $z=z_R$, which are considered as reference signals ($z_R=z_F$ for the direct problem, $z_R=z_A$ for the inverse problem)

$$T_1 = \frac{\langle p_r(x,y,z_R,t)\,p(x,y,z_R,t)\rangle}{\sqrt{\langle p_r^2(x,y,z_R,t)\rangle\langle p^2(x,y,z_R,t)\rangle}}, \quad T_2 = \sqrt{\frac{\langle p^2(x,y,z_R,t)\rangle}{\langle p_r^2(x,y,z_R,t)\rangle}}. \quad (15)$$

$\langle\ \rangle$ is the mean value. $T_1$ is a correlation coefficient which is sensitive to the similarity between the shapes of the signals and thus between their phase difference. $T_2$ is the ratio between two root mean square values for characterizing the similarity of the amplitudes of both signals.

### 3.3 Results in the time-space domain

Figure 3 highlights the temporal pressure signals $p(0.75\text{m}, 0.75\text{m}, 0.1575\text{m}, t)$ radiated in $P_2$ in the left, and $p(0.375\text{m}, 0.625\text{m}, 0.05\text{m}, t)$ back-propagated in $M_4$ in the right (see figure 1 for the locations). The pressure signal in $P_2$ is provided by the method (K) based on numerical Kaiser filtering while the Fourier method (F) is used to compute the impulse response to be inverted for resolving the inverse problem in $M_4$. The indicators $T_1$ and $T_2$ are given in figure 3. The study is also carried out in locations $P_1$, $P_2$, $P_3$, $P_4$, and $M_1$, $M_2$, $M_3$, $M_4$ for each method presented before.

Table 1. Indicators $T_1$, $T_2$ for time-space signal reconstruction in locations specified in figure 1

P1

| | D | A | C | K | F |
|---|---|---|---|---|---|
| T1 | 0.963 | 0.978 | 0.975 | 0.977 | 0.979 |
| T2 | 1.05 | 1.051 | 1.061 | 1.058 | 1.07 |

P2

| | D | A | C | K | F |
|---|---|---|---|---|---|
| T1 | 0.936 | 0.958 | 0.92 | 0.961 | 0.963 |
| T2 | 0.839 | 1.054 | 0.951 | 1.009 | 0.971 |

M1

| | D | A | C | K | F | I |
|---|---|---|---|---|---|---|
| T1 | 0.974 | 0.837 | 0.891 | 0.643 | 0.852 | 0.711 |
| T2 | 0.881 | 2.353 | 0.657 | 1.678 | 0.818 | 14.292 |

M2

| | D | A | C | K | F | I |
|---|---|---|---|---|---|---|
| T1 | 0.993 | 0.758 | 0.9 | 0.636 | 0.868 | 0.888 |
| T2 | 1.034 | 1.394 | 1.005 | 0.708 | 0.837 | 10.749 |

P3

| | D | A | C | K | F |
|---|---|---|---|---|---|
| T1 | 0.895 | 0.933 | 0.89 | 0.935 | 0.945 |
| T2 | 0.755 | 1.021 | 0.887 | 1.003 | 0.979 |

P4

| | D | A | C | K | F |
|---|---|---|---|---|---|
| T1 | 0.994 | 0.989 | 0.99 | 0.989 | 0.991 |
| T2 | 1.017 | 1.056 | 1.043 | 1.063 | 1.063 |

M3

| | D | A | C | K | F | I |
|---|---|---|---|---|---|---|
| T1 | 0.998 | 0.759 | 0.933 | 0.663 | 0.858 | 0.967 |
| T2 | 0.881 | 2.706 | 0.797 | 0.762 | 0.693 | 7.205 |

M4

| | D | A | C | K | F | I |
|---|---|---|---|---|---|---|
| T1 | 0.989 | 0.59 | 0.943 | 0.843 | 0.833 | 0.602 |
| T2 | 0.957 | 8.478 | 1.119 | 2.212 | 1.164 | 17.305 |

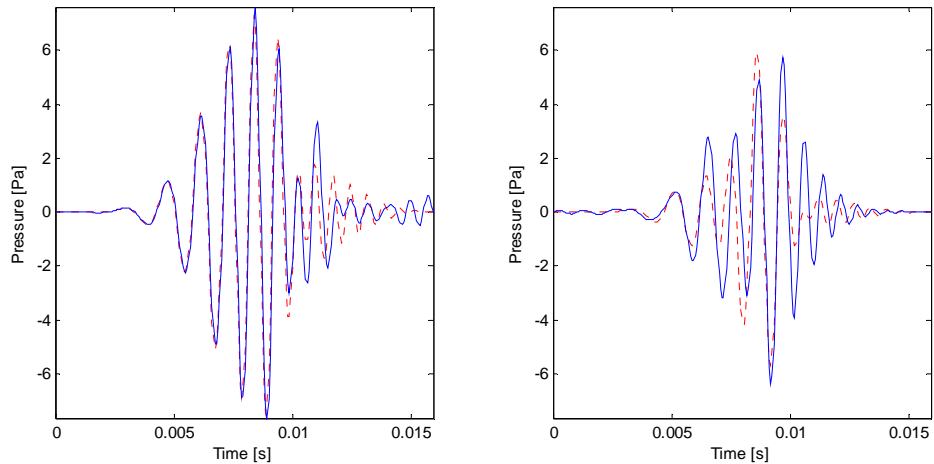

Figure 3. Reconstructed temporal signals versus reference signals (dotted line) in the time-space domain: the location is $P_2$ in the left (direct problem) with $T_1 = 0.961$ and $T_2 =1.009$. The location is $M_4$ in the right (inverse problem) with $T_1 = 0.833$ and $T_2 =1.164$.

## 4. CONCLUSIONS

Some methods to implement Real-Time Nearfield Acoustic Holography have been presented. For the radiation direct problem, the use of a low-pass Kaiser-Bessel filter or an inverse Fourier transform of the analytical frequency response leads to the most operational impulse response in the time-wavenumber domain. For the holography inverse problem, the best results are obtained by inverting using Wiener adaptive filtering the latter impulse response or that obtained by upsampling and Chebyshev low-pass filtering. However, some distortions in the reconstructed time pressure signals demonstrate that it would be necessary to improve the inverting process.